# Structural, magnetic and transport properties in the Cu-doped manganites $La_{0.85}Te_{0.15}Mn_{1-x}Cu_xO_3$ ( $0 \leq x \leq 0.20$ )

J. Yang[1], W. H. Song[1], Y. Q. Ma[1], R. L. Zhang[1], B. C. Zhao[1], Z.G. Sheng[1], G. H. Zheng[1], J. M. Dai[1], and Y. P. Sun*[1,2]

[1]*Key Laboratory of Materials Physics, Institute of Solid State Physics, Chinese Academy of Sciences, Hefei, 230031, P. R. China*

[2]*National Laboratory of Solid State Microstructures, Nanjing University, Nanjing 210008, P. R. China*

The effect of Cu-doping at Mn-site on structural, magnetic and transport properties in electron-doped manganites $La_{0.85}Te_{0.15}Mn_{1-x}Cu_xO_3$ ($0 \leq x \leq 0.20$) has been investigated. Based on the analysis of structural parameter variations, the valence state of the Cu ion in Cu-doped manganites is suggested to be +2. All samples undergo the paramagnetic-ferromagnetic (PM-FM) phase transition. The Curie temperature $T_C$ decreases and the transition becomes broader with increasing Cu-doping level, in contrast, the magnetization magnitude of Cu-doping samples at low temperatures increase as $x \leq 0.15$. The insulator-metal (I-M) transition moves to lower temperatures with increasing Cu-doping content and disappears as x > 0.1. In addition, the higher temperature resistivity ρ peak in double-peak-like ρ(T) curves observed in no Cu-doping sample is completely suppressed as Cu-doping level $x$ = 0.1 and ρ(T) curve only shows single I-M transition at the low temperature well below $T_C$. The results are discussed according to the change of magnetic exchange interaction caused by Cu-doping.

**PACS numbers:** 75.30.Kz  75.47.Lx  71.30.+h

The hole-doped magnanites $Ln_{1-x}A_xMnO_3$ (Ln =La-Tb, and A = Ca, Sr, Ba, Pb, etc.) have attracted much renewed attention due to the discovery of colossal magnetoresistance (CMR).[1] Many theories have been proposed to explain the mechanism about CMR such as double exchange (DE)[2], polaronic effects[3] and phase separation combined with percolation.[4] These theories suggest that the mixed valence of $Mn^{3+}/Mn^{4+}$ is a key component for understanding the CMR effect and the transition from the paramagnetic (PM) insulator-ferromagnetic (FM) metal. Recently, many researches have been focused on electron-doped compounds such as $La_{1-x}Ce_xMnO_3$,[5-7] $La_{1-x}Zr_xMnO_3$,[8] and $La_{1-x}Te_xMnO_3$[9-11] due to the potential applications in spintronics. These investigations also suggest that the CMR behavior probably occur in the mixed-valent state of $Mn^{2+}/Mn^{3+}$. The basic physics in terms of Hund's rule coupling between $e_g$ electrons and $t_{2g}$ core electrons and Jahn-Teller (JT) effect due to $Mn^{3+}$ JT ions can operate in the electron-doped manganites as well. The quasi-static ferromagnetism between $Mn^{3+}$-O-$Mn^{3+}$ possibly caused by the vibronic-induced JT effect is also possible.[12-13]

To study the substitution of Mn sites by transitional elements for clarifying the mechanism of CMR is very important because of the crucial role of Mn ions in the CMR materials. The Mn-site doping is an effective way to modify the crucial Mn-O-Mn network and in turn remarkably affects their intrinsic physical properties. So far, for hole-doped manganites, Mn-site doping effects have been extensively investigated.[14-19] Among the investigation of all Mn-site doping, the Cu-doping at Mn-site is more interesting because of its special nature of Cu perovskite compounds, i.e., its high temperature superconductivity of cuprates. However, many controversial results have been reported concerning the influence of Cu-doping on the properties of polycrystalline manganites. Ghosh et al. studied the effect of transitional element doping (TE = Fe, Co, Ni, Cr, Cu, Zn) in $La_{0.7}Ca_{0.3}MnO_3$ and found that the doping content of 5% Cu and Zn both lead to remarkably drop in resistance in the system.[14] The doped Cu in polycrystalline $La_{0.825}Sr_{0.175}Mn_{1-x}Cu_xO_3$ $(0 \leq x \leq 0.20)$ giving rise to the remarkable increase of the resistance has been found, which is attributed to the induced antiferromagnetic (AFM) superexchange interaction and the long range FM order becoming weaker.[15] Whereas in Cu heavily doped polycrystalline samples, two different kinds of conduction mechanism has been found in the doping range of $0.20 \leq x \leq 0.32$ and $0.33 \leq x \leq 0.40$, respectively.[16] In addition, the valence state of the doping Cu ion is a dispute. On the whole, at present, most of the investigations have been focused on Cu-doping at Mn-site in hole-doped manganites. In contrast, there are few reports on Cu-doping effect at Mn-site in electron-doped manganites. In order to comprehensively understand the origin of the effect of Cu-doping at Mn-site of electron-doped manganites, we carefully investigate the effect of Cu-doping at Mn-site on structural, magnetic and electrical transport properties in the electron-doped manganites $La_{0.85}Te_{0.15}Mn_{1-x}Cu_xO_3$ ($0 \leq x \leq 0.2$).

Polycrystalline $La_{0.85}Te_{0.15}Mn_{1-x}Cu_xO_3$ $(0 \leq x \leq 0.20)$ samples were prepared through the conventional solid-state reaction method in air. Appropriate proportions of high-purity $La_2O_3$, $TeO_2$, $MnO_2$ and $CuO$ powders were thoroughly mixed according to the desired stoichiometry, and then prefired at 700°C for 24h. The powders obtained were ground, pelletized, and sintered at 1000°C for 24h with three intermediate grinding, and finally, the furnace was cooled down to room temperature. The crystal structures were examined by x-ray diffractometer using Cu $K_\alpha$ radiation at room temperature. The magnetic measurements were carried out with a Quantum Design superconducting quantum interference device (SQUID) MPMS system ($2 \leq T \leq 400$ K, $0 \leq H \leq 5$ T). The resistance was measured by the standard four-probe method from 25 to 300 K.

Fig.1 shows XRD patterns of $La_{0.85}Te_{0.15}Mn_{1-x}Cu_xO_3$ $(0 \leq x \leq 0.20)$ samples. The powder x-ray diffraction at room temperature shows that all samples are single phase with no detectable secondary phases and the samples had a rhombohedral lattice with the space group $R\bar{3}C$. The structural parameters can be obtained by fitting the experimental spectra using the standard Rietveld technique.[20] The obtained structural parameters are listed in Table I. No structural transition has been observed and the lattice parameters for $La_{0.85}Te_{0.15}Mn_{1-x}Cu_xO_3$ $(0 \leq x \leq 0.20)$ samples vary monotonically with increasing Cu content. The Mn-O-Mn bond angle decreases with increasing Cu-doping level, whereas the Mn-O bond length increases which displays the inverse correlation to the variation in the Mn-O-Mn bond angle. To describe the ion match between A and B site ions in perovskite structure compounds, a geometrical quantity, i.e., tolerance factor $t$ is usually introduced and $t$ is defined as $t = (r_A + r_O)/\sqrt{2}(r_B + r_O)$, where $r_i$ ($i$=A, B, or O) represent the average ionic size of each element. It is well known that some internal stress will be introduced into the structure when there is a size mismatch between atom A and B sites. Internal stress can be partially relieved by structure distortion. The cubic structure is distorted either by atom B moving off center in its octahedra or by the cage collapsing by rotation of the $BO_6$ octahedra. For the present studied system, the distortion of $MnO_6$ octahedra arises from partial replacement of Mn by Cu because of the fixed average ionic radius of A-site. It is well known that the departure from the average Mn radius $R(Mn_{av})$ (defined as the average over the radii for $Mn^{3+}$ and $Mn^{4+}$ ions in the ratio (1-x): x in $Ln_{1-x}A_xMnO_3$ compounds) at the dopant site would subject the neighboring Mn-O bonds to a centric push or pull.[14] When a bigger ion at the Mn site should compress the neighboring Mn-O bonds and result in decreasing the Mn-O-Mn bond angle. Similarly, we can calculate $R(Mn_{av})$ (defined as the average radii for $Mn^{3+}$ and $Mn^{2+}$ ions in the ratio 85: 15 in $La_{0.85}Te_{0.15}MnO_3$ compounds). Standard ionic radii [21] with values 0.83Å for $Mn^{2+}$, and 0.65Å for $Mn^{3+}$ in perovskite-structure manganites, respectively, are used to calculate $R(Mn_{av})$. The average radius of Mn ion $R(Mn_{av})$ is calculated to be 0.658 Å. Therefore, based on the monotonic varying of Mn-O bond length, lattice volume, and Mn-O-Mn bond angle with increasing Cu-doping level, we suggest that the doped Cu ions at Mn-site should be in the form of $Cu^{2+}$ ions because the radius of the $Cu^{2+}$ is 0.73 Å, i.e., $R(Cu^{2+}) > R(Mn_{av})$.

The temperature dependence of magnetization M of

$La_{0.85}Te_{0.15}Mn_{1-x}Cu_xO_3$ $(0 \leq x \leq 0.20)$ under both zero-field cooling (ZFC) and field cooling (FC) modes at H = 0.1 T are measured. Our measured samples may be considered as ellipsoid for which N=0.05 (SI units), and the applied field is parallel to the longest semi-axis of samples. So a uniform field can exist throughout the samples and the shape demagnetizing fields can be reduced as much as possible. The results are shown in Fig.2 for the samples with $x <0.15$. For samples with $x = 0.15$ and $x = 0.20$, M-T curves are plotted in the inset of Fig.2. The Curie temperature $T_C$ (defined as the one corresponding to the peak of $dM/dT$ in the M vs. T curve) are 239 K, 193 K, 169 K, 111 K and 109 K for $x$ = 0, 0.05, 0.10, 0.15 and 0.20, respectively. Obviously, the Curie temperature $T_C$ decreases monotonically with increasing Cu-doping level. We suggest that the $T_C$ reduction could be attributed to the combined effect of the partial destruction of $Mn^{2+}$-O-$Mn^{3+}$ DE interaction network and the weakening of DE interaction because of the decrease of the bandwidth and the mobility of $e_g$ electrons due to the increase of Mn-O bond length and the decrease of Mn-O-Mn bond angle caused by Cu-doping.

In addition, from Fig.2, a sharp FM to PM transition is observed for $x = 0$ and $x = 0.05$ samples. However, as $x > 0.05$, the temperature range of FM-PM phase transition becomes broader with increasing Cu-doping level, which implies the increase of magnetic inhomogeneity. The bifurcation begins to occur between the FC and ZFC curves at low temperature region for the samples with $0.10 \leq x \leq 0.20$ implying the appearance of magnetic frustration. With increasing $Cu^{2+}$ content, a bigger difference between M-T curves under FC and ZFC is observed. For samples with $x = 0.15$ and 0.20, an obvious spin-glass-like behavior appears, which is evidenced by a cusp in the ZFC curve at 40K and a distinctive separation of the FC and ZFC curves. Although the decrease of $T_C$ and broadening of PM-FM transition with increasing Cu-doping level, the magnetization magnitude of Cu-doping samples increases at low temperatures with increasing Cu-doping level as $x \leq 0.15$. The magnetization as a function of the applied magnetic field at 5K is shown in Fig.3, The magnetization reaches saturation at about 1T and keeps constant up to 5T for the samples with $x = 0$, 0.05 and 0.1, which is considered as a result of the rotation of the magnetic domain under external magnetic field, whereas for the samples with x=0.15 and 0.2, the M(H) dependences are unsaturated up to H=5T and they exhibit high slope. It is well known that the competition between FM phase and AFM phase would lead to the appearance of a spin-glass state. That is why a spin-glass-like behavior exists in the samples with $x = 0.15$ and 0.20. Additionally, similar to the result shown in Fig.2, the magnetization magnitude of Cu-doping samples also increases at low temperatures as H > 1 T for $x \leq 0.15$ compared with that of no Cu-doped sample. It may be related to the appearance of superexchange ferromagnetism (SFM) due to the obvious deviation of 180° of Mn-O-Mn bond angle. The SFM phase is in fact a FM phase with predominant superexchange ferromagnetic interaction. As Goodenough predicts, a Mn-O-Mn 180° superexchange interaction generally gives rise to AFM ordering while 90° superexchange interaction will lead to FM ordering.[22] This SFM has been suggested to explain ferromagnetism of $Tl_2Mn_2O_7$, in which there is no Mn mixed valence.[23] On the other hand, the opening of the new DE channel between $Mn^{3+}$-O-$Mn^{4+}$ is also possible because the substitution of $Cu^{2+}$ for $Mn^{3+}$ introduces $Mn^{4+}$ ion.

The temperature dependence of resistivity for $La_{0.85}Te_{0.15}Mn_{1-x}Cu_xO_3$ $(0 \leq x \leq 0.20)$ is shown in Fig.4. The experimental data are obtained at zero and applied field of 0.5 T for the samples with $x = 0$, 0.05 and 0.10 in the temperature range of 25-300K. Fig.4 shows that ρ increase considerably with increasing Cu-doping level. For no Cu-doped sample, it shows that there exists an I-M transition peak at $T_{P1}$ (=246 K) slightly higher than its $T_C$ (=239 K). In addition, there exists a shoulder at $T_{P2}$ (=223 K) below $T_C$. Double peaks ($T_{P1}$ = 196 and $T_{P2}$ = 176 K) shift to low temperatures for $x = 0.05$ sample. Compared with the $x = 0$ sample, I-M transition at $T_{P1}$ becomes weak and I-M transition at $T_{P2}$ becomes more obvious. It shows that the Cu-doping at Mn-site can substantially enhance the I-M transition at $T_{P2}$.

ρ(T) curve only exhibits single pronounced I-M transition at $T_P$ = 117 K for the sample with $x$ = 0.1, which lies in well below its $T_C$ value of 169 K. It shows that the I-M transition at $T_{P1}$ has been suppressed completely for $x \geq 0.1$. This variation of double ρ peak behavior is presumably related to an increase both of the height and width of tunnel barriers with increasing Cu-doping.[24] Moreover, Fig.4 manifests that ρ increases several orders of magnitude as $x$ > 0.05. For the samples with $x$ = 0.15 and 0.20, the resistance at low temperature is so high that the data are collected merely in a limit temperature range in order to avoid exceeding our measuring limit and ρ(T) displays the semiconducting behavior ($d\rho/dT < 0$) in the whole measurement temperature range. It means that the samples with $x$ >0.1 exhibit semiconducting behavior in both high-temperature PM phase and low-temperature FM phase. We suggest that the remarkable increase of ρ for Cu-doping samples originates from the combined effect of destruction of $Mn^{2+}$-O-$Mn^{3+}$ DE interaction network, the appearance of SFM insulating phase and the introduction of random Coulomb potential caused by the substitution of $Cu^{2+}$ for $Mn^{3+}$.

The temperature dependence of resistivity of the samples with $x$ = 0, 0.05 and 0.1 at a magnetic field of 0.5 T is also plotted in Fig.4. It shows that the resistivity of samples decreases under the applied magnetic field, $T_{P1}$ peak position shifts to a higher temperature and $T_{P2}$ peak position does not nearly change, which also means the origin of $T_{P2}$ peak is different from that of $T_{P1}$ peak. The magnetoresistance (MR) as a function of temperature based on the data shown in Fig.4 is plotted in Fig.5. Here the MR is defined as $\Delta\rho/\rho_H = (\rho_0 - \rho_H)/\rho_H$, where $\rho_0$ is the resistivity at zero field and $\rho_H$ is the resistivity under 0.5T. For the samples with $x$= 0 and 0.05, there are corresponding peaks in the vicinity of $T_{P1}$ on the MR curves, and then the MR values increase with decreasing temperature. The similar MR behavior has been usually observed in polycrystalline samples of hole-doped manganites which is considered to be related to grain boundaries.[25] For the sample with $x = 0.10$, no obvious MR peak close to $T_P$ is observed. However, there exist a slight enhanced MR at low temperatures.

The effect of Cu-doping on structural, magnetic and transport properties in electron-doped manganites $La_{0.85}Te_{0.15}Mn_{1-x}Cu_xO_3$ $(0 \leq x \leq 0.20)$ has been studied. The valence state of the Cu ion is determined to be +2. The Curie temperature $T_C$ of samples decreases with increasing Cu-doping level. The magnetization magnitude of Cu-doping samples at low temperatures increase as $x \leq 0.15$ although their $T_C$ is reduced, which is considered to be related to the appearance of SFM and a possible DE interaction of $Mn^{3+}$-O-$Mn^{4+}$ due to the introduction of $Mn^{4+}$ ion because of Cu-doping. The higher temperature resistivity ρ peak in double-peak-like ρ(T) curves observed in no Cu-doping sample is completely suppressed as Cu-doping level up to $x = 0.1$ and ρ(T) curve only shows single I-M transition at low temperatures. For the samples with $x > 0.1$, ρ(T) curves exhibit semiconducting behavior ($d\rho/dT < 0$) in both high-temperature PM phase and low-temperature FM phase, which is suggested to stem from the appearance of SFM insulating phase caused by Cu-doping.

This work was supported by the National Key Research under contract No.001CB610604, and the National Nature Science Foundation of China under contract No.10174085, Anhui Province NSF Grant No.03046201 and the Fundamental Bureau Chinese Academy of Sciences. The first author would like to thank Dr. K.Y. Wang for the aid in the preparation of the manuscript.

# Tables

TABLE I. Refined structural parameters of $La_{0.85}Te_{0.15}Mn_{1-x}Cu_xO_3 (0 \leq x \leq 0.20)$ at room temperature. The space group is $R\bar{3}C$.

| Parameter | x = 0 | x = 0.05 | x = 0.10 | x = 0.15 | x = 0.20 |
|---|---|---|---|---|---|
| a (Å) | 5.525 | 5.530 | 5.531 | 5.535 | 5.538 |
| c (Å) | 13.355 | 13.348 | 13.348 | 13.346 | 13.343 |
| v (Å$^3$) | 353.010 | 353.273 | 353.589 | 354.049 | 354.340 |
| Mn-O (Å) | 1.9644 | 1.9690 | 1.9721 | 1.9849 | 1.9911 |
| Mn-O-Mn (°) | 163.83 | 162.10 | 161.28 | 157.49 | 155.92 |
| R$_p$ (%) | 8.03 | 6.23 | 9.29 | 6.74 | 6.73 |

# Figure captions

Fig.1. X-ray diffraction patterns for $La_{0.85}Te_{0.15}Mn_{1-x}Cu_xO_3$ ($x$ = 0, 0.05, 0.1, 0.15 and 0.2) samples.

Fig.2. Magnetization as a function of temperature for $La_{0.85}Te_{0.15}Mn_{1-x}Cu_xO_3$ ($x$ = 0, 0.05 and 0.1) measured at H = 0.1T under the field-cooled (FC) and zero-field-cooled (ZFC) modes that are denoted as the filled and open symbols, respectively. The inset show M (T) curves for the samples with x = 0.15 and 0.2.

Fig.3. Field dependence of the magnetization in $La_{0.85}Te_{0.15}Mn_{1-x}Cu_xO_3$ (x = 0, 0.05, 0.1, 0.15 and 0.2) at 5 K.

Fig.4. The temperature dependence of the resistivity of $La_{0.85}Te_{0.15}Mn_{1-x}Cu_xO_3$ ($x$ = 0, 0.05, 0.1, 0.15 and 0.2) samples at zero (solid lines) and 0.5T fields (dashed lines).

Fig.5. The temperature dependence of magnetoresistance (MR) ratio of $La_{0.85}Te_{0.15}Mn_{1-x}Cu_xO_3$ at 0.5 T field for the samples with $x$ = 0, 0.05 and 0.1.

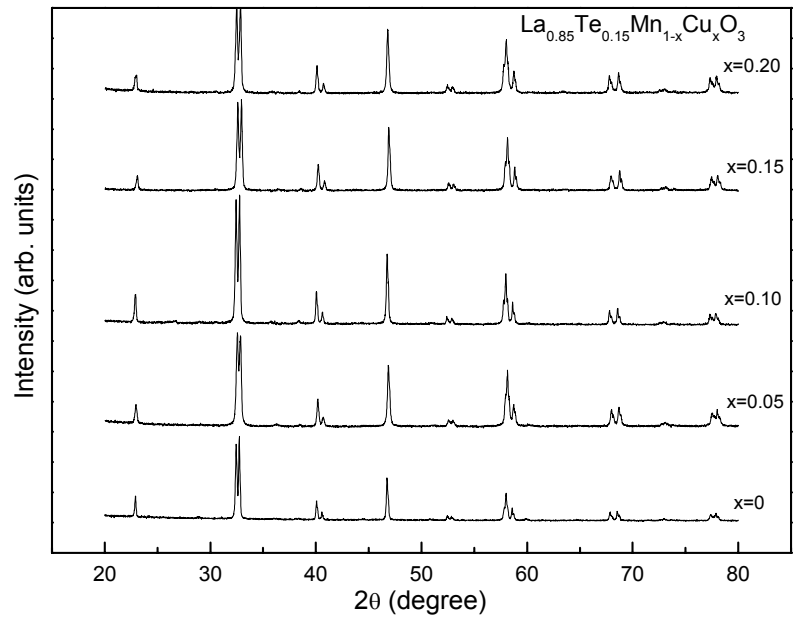

Fig.1. J. Yang et al.

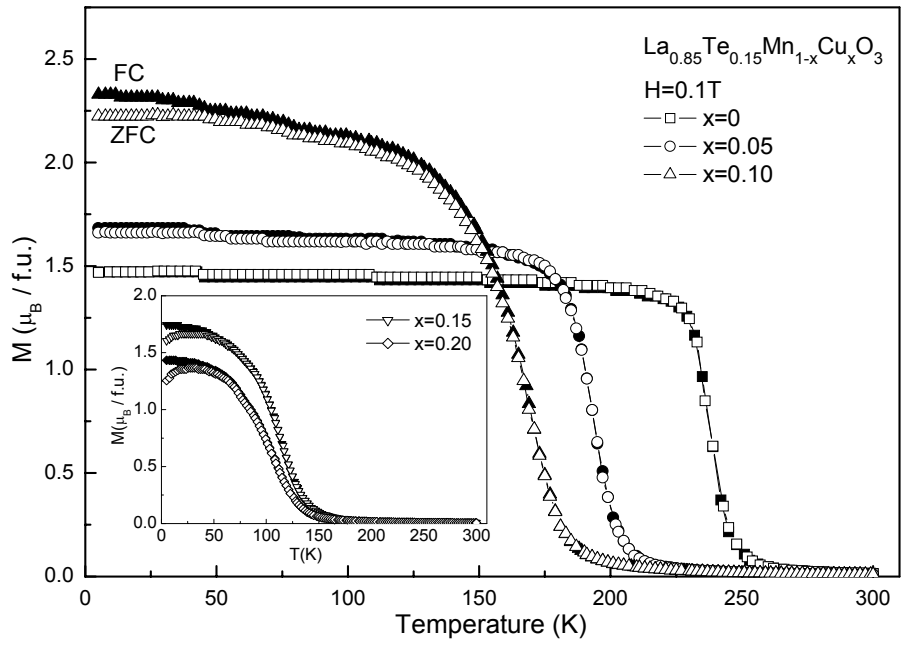

Fig.2. J. Yang et al.

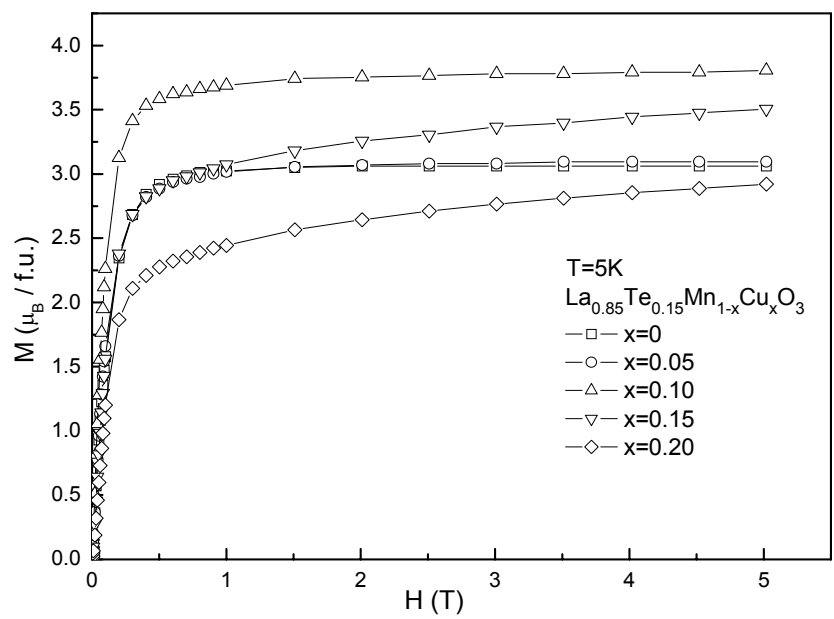

Fig.3. J. Yang et al.

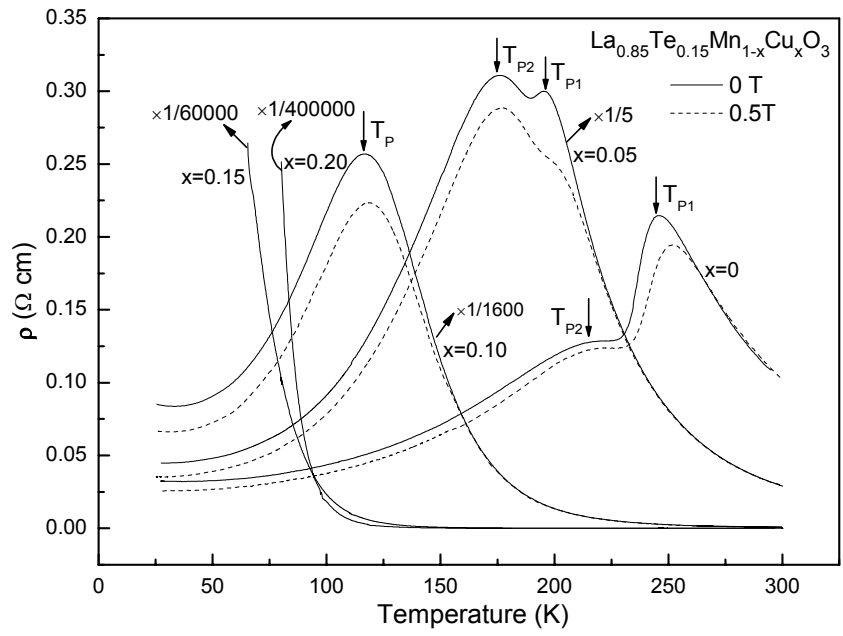

Fig.4. J. Yang et al.

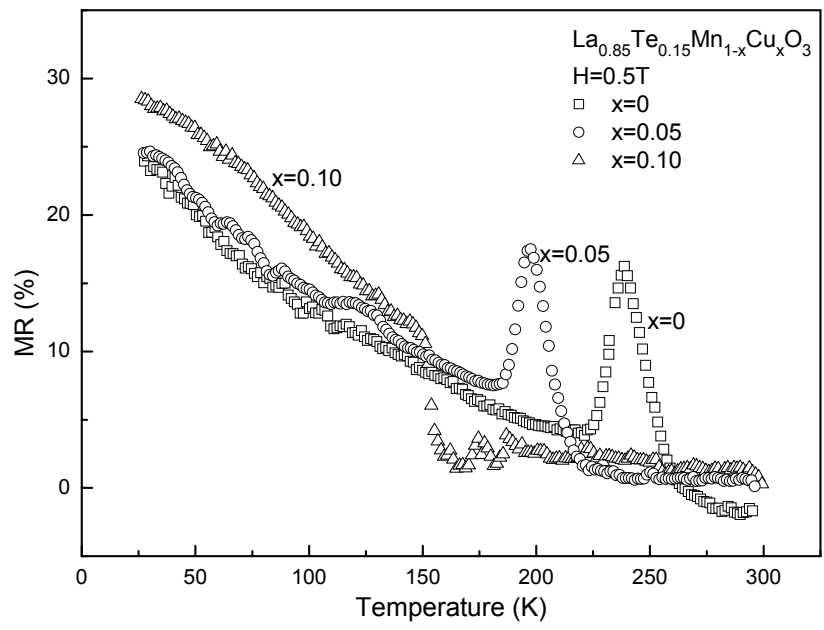

Fig.5. J. Yang et al.